\documentclass{ws-procs9x6}

\usepackage{graphicx}

\newcommand{\ppara}{\ensuremath{p_\parallel}}

\newcommand{\Ppara}{\ensuremath{P_\parallel}}
\newcommand{\qpara}{\ensuremath{q_\parallel}}
\newcommand{\qperp}{\ensuremath{q_\perp}}

\newcommand{\gammapara}{\ensuremath{\gamma_\parallel}}
\begin{document}

\title{Chiral Symmetry Breaking and Stability of the Magnetized Vacuum}

\author{Shang-Yung Wang}

\address{Department of Physics, Tamkang University,\\ Tamsui, Taipei 25137, Taiwan\\
E-mail: sywang@mail.tku.edu.tw}

\begin{abstract}
The recent claim that there exists in QED a maximum magnetic field of $10^{42}$~G, above which the magnetized vacuum becomes unstable with respect to the so-called ``positronium collapse'' is critically examined and unequivocally refuted.
\end{abstract}

%\keywords{Chiral symmetry breaking; Magnetic field; Schwinger-Dyson equations, Bethe-Salpeter equation}

\vspace{0.25truein}

\bodymatter
It is an observational fact that we are living in a magnetized universe, with planets, stars and galaxies all being magnetized~\cite{Giovannini:2003yn}. The typical magnetic fields range from few $\mu$G in galaxies to few G on planets (such as the earth) and up to $10^{12}-10^{16}$~G on the surface of neutron stars and magnetars. A fundamental question to ask is whether there exists in nature a maximum magnetic field. As we will see shortly, the answer to this question requires a nonperturbative understanding of quantum field theory under the influence of a strong magnetic field.

Conventional wisdom suggests that the magnetic field strength is not constrained by known physics. However, a maximum value for the magnetic field in QED, $B_\mathrm{max}$, has recently been conjectured~\cite{Shabad:2006ci}:
\begin{equation}
B_\mathrm{max}=\frac{m^2}{4e}\,\exp\bigg(\frac{\pi^{3/2}}{\sqrt{\alpha}}+2C_\mathrm{E}\bigg)\simeq 10^{42}~\mathrm{G},\label{Bmax}
\end{equation}
where $m$ and $e$ are respectively the electron mass and the absolute value of its charge in the absence of external fields, $\alpha=e^2/4\pi$ is the fine structure constant and $C_\mathrm{E}\simeq 0.577$ is Euler's constant.
In obtaining \eqref{Bmax}, Shabad and Usov~\cite{Shabad:2006ci} considered a positronium (electron-positron bound state) placed in a strong magnetic field, and found that the magnetic field significantly enhances the Coulomb attraction between the constituent electron and positron. The Coulomb attraction becomes stronger and stronger until the electron and positron fall onto each other at the maximum magnetic field of $10^{42}$~G. This phenomenon is referred to by these authors as ``positronium collapse'' and could be a signal for possible vacuum instability~\cite{Shabad:2006ci}.

While the existence of a maximum magnetic field in QED appears novel and is potentially of fundamental importance, it is in fact in contradiction with many of the well-established results in QED in a strong magnetic field~\cite{Gusynin:1995gt,Gusynin:1998zq,Leung:2005yq,Wang:2007sn}. In particular, magnetic catalysis of chiral symmetry breaking has long been known as a universal phenomenon. A strong magnetic field acts as a catalyst for chiral symmetry breaking, leading to the generation of a dynamical fermion mass even at the weakest attractive interaction between fermions. The hallmark of this phenomenon is the dimensional reduction from $(3+1)$ to $(1+1)$ in the dynamics of fermion pairing in a strong magnetic field when the lowest Landau level (LLL) plays the dominant role. The massless fermion-antifermion bound state is the Nambu-Goldstone (NG) boson for spontaneously broken chiral symmetry~\cite{Gusynin:1995gt}. The phenomenon of magnetic catalysis is universal in that chiral symmetry is broken in arbitrarily strong magnetic fields and for any number of fermion flavors~\cite{Gusynin:1995gt,Gusynin:1998zq,Leung:2005yq}.

A powerful tool to study the electron-position bound states directly from QED is the Bethe-Salpeter (BS) equation. For the problem at hand, however, in order to consistently incorporate the effective electron mass and the screening effect in a strong magnetic field, the electron and photon propagators that enter the BS equation must be obtained by solving the corresponding Schwinger-Dyson (SD) equations in the same truncation. This is tantamount to solving the truncated SD and BS equations simultaneously, an important point that has gone unnoticed in Ref.~\refcite{Shabad:2006ci}.

It has recently been proved~\cite{Leung:2005yq,Wang:2007sn} that in QED (both massless and massive) in a strong magnetic field the bare vertex approximation (BVA), in which the vertex corrections are completely ignored, is a consistent truncation of the SD equations within the lowest Landau level approximation (LLLA). In particular, it can be shown that the truncated vacuum polarization is transverse and the dynamical fermion mass, obtained as the solution of the truncated fermion SD equation evaluated on the fermion mass shell, is manifestly gauge independent. Thus, for consistency with the results obtained in Refs.~\refcite{Leung:2005yq,Wang:2007sn}, the BS equation for the positronium has to be truncated in the BVA within the LLLA.

\begin{figure}[t]
\begin{center}
\includegraphics[height=0.5truein,keepaspectratio=true,clip=true]{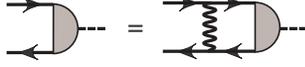}
\end{center}
\caption{BS equation for the positronium in the BVA. External lines are amputated.}
\label{fig:BSBVA}
\end{figure}

We choose the constant external magnetic field of strength $B>0$ in the $x_3$-direction. The corresponding vector potential is given by $A^\mathrm{ext}_\mu=(0,0,Bx_1,0)$. Note that the motion of the LLL electron and positron is restricted in directions perpendicular to the magnetic field, hence so is the motion of the bound state of the LLL electron and positron. We will henceforth refer to the latter as the LLL positronium. In the BVA within the LLLA, the BS equation for the LLL positronium in momentum space is given by~\cite{Leung:2009gw} (see Fig.~\ref{fig:BSBVA})
\begin{align}
\chi(\ppara;\Ppara)\,\Delta=&-ie^2\int_q\,\exp\biggl(-\frac{q_\perp^2}{2eB}\biggr)\,
\mathcal{D}_{\mu\nu}(q)\,\Delta\,\gammapara^\mu\,\Delta\,\frac{1}{\gammapara\cdot\ppara'+m_\ast}\,\chi(\ppara';\Ppara)\nonumber\\
&\times\frac{1}{\gammapara\cdot\ppara'+m_\ast}\,\Delta\,\gammapara^\nu\,\Delta,\label{chi2}
\end{align}
where $\chi(\ppara;\Ppara)$ is the amputated BS amplitude, $\ppara^\mu=(p^0,p^3)$ is the momentum of the LLL electron, $\Ppara^\mu$ is the momentum of the LLL positronium, $m_\ast$ is the effective electron mass in a strong magnetic field, $\ppara'=\ppara-\qpara$, $\qperp^2=q_1^2+q_2^2$ and $\int_q=\int d^4q/(2\pi)^4$. In \eqref{chi2},  $\Delta=(1+i\gamma^1\gamma^2)/2$ is the projector onto the electron (positron) states with the spin polarized along (opposite to) the external magnetic field and $\mathcal{D}_{\mu\nu}(q)$ is the full photon propagator in the external magnetic field. The latter in covariant gauges is given by~\cite{Gusynin:1998zq,Leung:2005yq,Wang:2007sn}
\begin{equation}
\mathcal{D}^{\mu\nu}(q)=\frac{1}{q^2+\Pi(\qpara^2,\qperp^2)}
\biggl(g_\parallel^{\mu\nu}-\frac{q^\mu_\parallel
q^\nu_\parallel}{\qpara^2}\biggr)+\frac{g_\perp^{\mu\nu}}{q^2}+\frac{q^\mu_\parallel q^\nu_\parallel}{q^2\qpara^2}
+(\xi-1)\frac{1}{q^2}\frac{q^\mu q^\nu}{q^2},\label{D}
\end{equation}
where $\Pi(\qpara^2,\qperp^2)$ is the polarization function (see Refs.~\refcite{Gusynin:1998zq,Leung:2005yq,Wang:2007sn} for an explicit expression).

Because the LLL electron and positron always have their spins polarized in opposite directions along the external magnetic field, the LLL positronium in its ground state is in fact a parapositronium, i.e., a pseudoscalar state. This, together with symmetry arguments, implies that the amputated BS amplitude $\chi(\ppara;\Ppara)$ takes the form $\chi(\ppara;\Ppara)=A(\ppara,\Ppara)\gamma^5$, where $A(\ppara,\Ppara)$ is a scalar function of $\ppara^2$ and $\Ppara^2$. Then, \eqref{chi2} simplifies to
\begin{align}
A(\ppara,\Ppara)=&-ie^2\int_q\,\exp\biggl(-\frac{q_\perp^2}{2eB}\biggr)
\mathcal{D}_{\mu\nu}(q)A(\ppara',\Ppara)\gammapara^\mu\,\frac{1}{\gammapara\cdot\ppara'+m_\ast}
\nonumber\\&\times
\frac{1}{\gammapara\cdot\ppara'+m_\ast}\,\gammapara^\nu.\label{A}
\end{align}
When evaluated on the respective particle mass shells, $\ppara^2=-m_\ast^2$ and $\Ppara^2=-M^2$, and supplemented with the effective electron mass $m_\ast$ that is obtained as the solution of the on-shell SD equations~\cite{Leung:2005yq,Wang:2007sn}, the BS equation \eqref{A} determines the mass of the LLL positronium, $M$.

\begin{figure}[t]
\begin{center}
\includegraphics[height=0.65truein,keepaspectratio=true,clip=true]{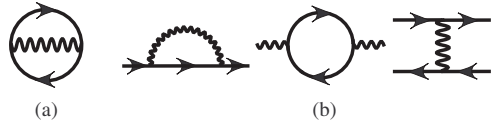}
\end{center}
\caption{(a) Contribution to the 2PI effective action at two-loop order. (b) The electron self-energy, vacuum polarization and electron-positron interaction kernel generated thereby. External lines are amputated.}
\label{fig:2PI}
\end{figure}

A detailed analysis based on the Ward-Takahashi identity in the BVA within the LLLA reveals that contrary to the gauge independence of the on-shell SD equations~\cite{Leung:2005yq,Wang:2007sn}, the on-shell BS equation truncated in the BVA is inevitably gauge dependent. We now argue that in the BVA the on-shell BS equation has a controlled gauge dependence, thanks to a direct correspondence~\cite{Leung:2005yq} between the SD equations truncated in the BVA and the 2PI effective action  truncated at the lowest nontrivial (two-loop) order in the loop expansion. Let $\Gamma_2$ denotes the sum of all 2PI skeleton vacuum diagrams with \emph{bare} vertex and \emph{full} LLL electron and photon propagators. The contribution to $\Gamma_2$ at two-loop order is depicted diagrammatically in Fig.~\ref{fig:2PI}(a). The key point of the argument is to note that the direct correspondence can be generalized to include the BS equation truncated in the BVA. This is because, as shown in Fig.~\ref{fig:2PI}(b), the corresponding electron self-energy, vacuum polarization and electron-positron interaction kernel that enter the SD and BS equations in the BVA are the same as those generated by $\Gamma_2$. The argument is completed with the fact~\cite{Arrizabalaga:2002hn} that the truncated 2PI effective action evaluated at its stationary point has a controlled gauge dependence, i.e., the explicit gauge dependent terms always appear at higher order. Moreover, since the transverse components in $\mathcal{D}^{\mu\nu}(q)$ decouple and the gauge dependent contribution in \eqref{A} arises from the longitudinal components in $\mathcal{D}^{\mu\nu}(q)$ proportional to $q^\mu q^\nu /q^2$, we conclude that in the BVA and at the order of truncation only the first term in $\mathcal{D}^{\mu\nu}(q)$ proportional to $g_\parallel^{\mu\nu}$ contributes to the on-shell BS equation.

To prove the nonexistence of a maximum magnetic field in QED, we only need to consider the on-shell BS equation in an asymptotically strong magnetic field. For $B\gg B_0\equiv m^2/e$, the explicit breaking of chiral symmetry associated with the perturbative electron mass $m$ can be neglected. As per the fact~\cite{Wang:2007sn} that $m_\ast\approx m_\mathrm{dyn}$ as $B\to\infty$ and the pseudo NG boson nature of the positronium that $M\to 0$ as $B\to\infty$, the BS equation \eqref{A} is evaluated on the mass shells $\ppara^2=-m_\mathrm{dyn}^2$ and $\Ppara^2=0$. Moreover, because of the strong screening effect in a strong magnetic field~\cite{Gusynin:1998zq,Leung:2005yq,Wang:2007sn}, the integral in \eqref{A} is dominated by contributions from the region with $m_\mathrm{dyn}^2\lesssim |\qpara^2|\ll eB$. Thus, $A(\ppara',\Ppara)$ in the integrand can be approximated by $A(\ppara,\Ppara)$. We find that in the limit $B\to\infty$ the on-shell BS equation reduces to
\begin{equation}
1=-2ie^2\int_q\,\frac{\exp(-q_\perp^2/2eB)}{q^2+\Pi(\qpara^2,\qperp^2)}
\frac{1}{(p-q)_\parallel^2+m_\mathrm{dyn}^2}\bigg|_{\ppara^2=-m_\mathrm{dyn}^2}.\label{gap}
\end{equation}
The proof is completed by noting that \eqref{gap} is the same as the on-shell SD equation obtained
in the BVA within the LLLA~\cite{Leung:2005yq} that reliably determines the dynamically generated fermion mass, $m_\mathrm{dyn}$, in massless QED (see (4.20) in the first paper of Ref.~\refcite{Leung:2005yq}). This also serves to justify \emph{a posteriori} the controlled gauge dependence of the on-shell BS equation.

In conclusion, the positronium is unambiguously identified as the (pseudo) Nambu-Goldstone boson for spontaneous (explicit) chiral symmetry breaking in massless (massive) QED in a strong magnetic field. It is shown that the phenomenon of positronium collapse conjectured by Shabad and Usov never takes place. Consequently, there does not exist a maximum magnetic field in QED and the magnetized vacuum is stable for all values of the magnetic field.

\vspace{0.1truein}
\emph{Acknowledgements}. The author would like to thank the organizers of the conference for their invitation and hospitality. This work was supported in part by the National Science Council of Taiwan under grant 96-2112-M-032-005-MY3.


\begin{thebibliography}{9}

%\cite{Giovannini:2003yn}
\bibitem{Giovannini:2003yn}
M. Giovannini,
%``The magnetized universe,''
Int. J. Mod. Phys.  D \textbf{13}, 391 (2004).
%%CITATION = IMPAE,D13,391;%%

%\cite{Shabad:2006ci}
\bibitem{Shabad:2006ci}
A. E. Shabad and V. V. Usov,
%``Positronium collapse and the maximum magnetic field in pure QED,''
Phys. Rev. Lett. \textbf{96}, 180401 (2006);
%[arXiv:hep-th/0605020].
%%CITATION = PRLTA,96,180401;%%
%
%\cite{Shabad:2006gf}
%\bibitem{Shabad:2006gf}
%A. E. Shabad and V. V. Usov,
%``Bethe-Salpeter approach for relativistic positronium in a strong  magnetic
%field,''
Phys. Rev. D \textbf{73}, 125021 (2006).
%[arXiv:hep-th/0603070].
%%CITATION = PHRVA,D73,125021;%%

%\cite{Gusynin:1995gt}
\bibitem{Gusynin:1995gt}
V. P. Gusynin, V. A. Miransky, and I. A. Shovkovy,
%``Dynamical chiral symmetry breaking by a magnetic field in QED,''
Phys. Rev. D \textbf{52}, 4747 (1995);
%[arXiv:hep-ph/9501304].
%%CITATION = HEP-PH 9501304;%%
%
%\cite{Gusynin:1995nb}
%\bibitem{Gusynin:1995nb}
%V. P. Gusynin, V. A. Miransky and I. A. Shovkovy,
%``Dimensional reduction and catalysis of dynamical symmetry breaking by a
% magnetic field,''
Nucl. Phys. \textbf{B462}, 249 (1996);
%[arXiv:hep-ph/9509320].
%%CITATION = HEP-PH 9509320;%%
%
%\cite{Lee:1997zj}
%\bibitem{Lee:1997zj}
D.-S. Lee, C. N. Leung, and Y. J. Ng,
%``Chiral symmetry breaking in a uniform external magnetic field,''
Phys. Rev. D \textbf{55}, 6504 (1997).
%[arXiv:hep-th/9701172].
%%CITATION = HEP-TH 9701172;%%

%\cite{Gusynin:1998zq}
\bibitem{Gusynin:1998zq}
V. P. Gusynin, V. A. Miransky, and I. A. Shovkovy,
%``Dynamical chiral symmetry breaking in QED in a magnetic field: Toward exact
%results,''
Phys. Rev. Lett. \textbf{83}, 1291 (1999);
%[arXiv:hep-th/9811079].
%%CITATION = HEP-TH 9811079;%%
%
%\cite{Gusynin:1999pq}
%\bibitem{Gusynin:1999pq}
%V. P. Gusynin, V. A. Miransky and I. A. Shovkovy,
%``Theory of the magnetic catalysis of chiral symmetry breaking in QED,''
Nucl. Phys. \textbf{B563}, 361 (1999);
%[arXiv:hep-ph/9908320].
%%CITATION = HEP-PH 9908320;%%
%
%\cite{Gusynin:2003dz}
%\bibitem{Gusynin:2003dz}
%V. P. Gusynin, V. A. Miransky and I. A. Shovkovy,
%``Large N dynamics in QED in a magnetic field,''
Phys. Rev. D \textbf{67}, 107703 (2003).
%[arXiv:hep-ph/0304059].
%%CITATION = HEP-PH 0304059;%%

%\cite{Leung:2005yq}
\bibitem{Leung:2005yq}
C. N. Leung and S.-Y. Wang,
%``Gauge independent approach to chiral symmetry breaking in a strong magnetic
%field,''
Nucl. Phys. \textbf{B747}, 266 (2006);
%%CITATION = HEP-PH 0510066;%%
%
%\cite{Leung:2005xz}
%\bibitem{Leung:2005xz}
%C. N. Leung and S.-Y. Wang,
%``Gauge independence and chiral symmetry breaking in a strong magnetic
%field,''
Ann. Phys. (N.Y.) \textbf{322}, 701 (2007).
%[arXiv:hep-ph/0503298]
%%CITATION = APNYA,322,701;%%

%\cite{Wang:2007sn}
\bibitem{Wang:2007sn}
S.-Y. Wang,
%``Dynamical Electron Mass in a Strong Magnetic Field,''
Phys. Rev. D \textbf{77}, 025031 (2008).
%[arXiv:0709.4427 [hep-ph]].
%%CITATION = PHRVA,D77,025031;%%

%\cite{Leung:2009gw}
\bibitem{Leung:2009gw}
C. N. Leung and S.-Y. Wang,
%``Is There a Maximum Magnetic Field in QED?,''
Phys. Lett. B \textbf{674}, 344 (2009).
%%CITATION = PHLTA,B674,344;%%

%\cite{Arrizabalaga:2002hn}
\bibitem{Arrizabalaga:2002hn}
A. Arrizabalaga and J. Smit,
%``Gauge-fixing dependence of Phi-derivable approximations,''
Phys. Rev. D \textbf{66}, 065014 (2002);
%[arXiv:hep-ph/0207044].
%%CITATION = HEP-PH 0207044;%%
%
%\cite{Carrington:2003ut}
%\bibitem{Carrington:2003ut}
M. E. Carrington, G. Kunstatter, and H. Zaraket,
%``2PI effective action and gauge invariance problems,''
Eur. Phys. J. C \textbf{42}, 253 (2005).
%%CITATION = HEP-PH 0309084;%%

\end{thebibliography}
\end{document}